\documentclass{article}

\usepackage{arxiv}

\usepackage[utf8]{inputenc} 
\usepackage[T1]{fontenc}    
\usepackage{hyperref}       
\usepackage{url}            
\usepackage{booktabs}       
\usepackage{amsfonts}       
\usepackage{nicefrac}       
\usepackage{microtype}      
\usepackage{lipsum}
\usepackage{multirow}
\usepackage{booktabs}
\usepackage{tabularx}
\usepackage{amsmath}
\usepackage[export]{adjustbox}
\usepackage{subcaption}
\usepackage{graphicx}
\graphicspath{ {./figures/} }

\title{Song Hit Prediction: Predicting Billboard Hits Using Spotify Data}

\author{
  Kai~Middlebrook\thanks{learn more at musicai.io} \\
  Department of Math \& Statistics \\
  University of San Francisco \\
  San Francisco, CA 94117 \\
  \texttt{kai@musicai.io} \\
   \And
 Kian Sheik \thanks{learn more at musicai.io}\\
  Department of Math \& Statistics \\
  University of San Francisco \\
  San Francisco, CA 94117 \\fdfdfdfd
  \texttt{kian@musicai.io} \\
}

\begin{document}
\maketitle

\begin{abstract}
In this work, we attempt to solve the Hit Song Science
problem, which aims to predict which songs will become chart-topping hits. We constructed a dataset with approximately 1.8 million hit and non-hit songs and extracted their audio
features using the Spotify Web API. We test four models on our dataset. Our best model was random forest, which was able to predict Billboard song success with 88\% accuracy. 
\end{abstract}


\section{Introduction}
Hit song science (HSS) aims to predict whether a given song will become a chart-topping hit.  The underlying assumption in HSS is that hit songs are similar with respect to their features. Hit Song Science is  an active research topic in Music Information Retrieval (MIR). 

Hit prediction is useful to musicians, labels, and music vendors because popular songs generate larger revenues and allow artists to share their message with a broad audience. For example, if a label would like to increase profits, they may choose to invest their limited resource (ad campaigns, studio equipment, etc.) on tracks that are likely to become popular. On the other hand, if an artist wants to embody an aesthetic that is devoid of mainstream musical characteristics, they may choose to release tracks that are unlikely to become popular. We describe our approach to solve the HSS problem in the proceeding sections.

\section{Methods}
\label{sec:methods}
\subsection{Dataset and Features}
\begin{figure}[ht]
    \begin{subfigure}{0.5\textwidth}
        \includegraphics[width=7cm, left]{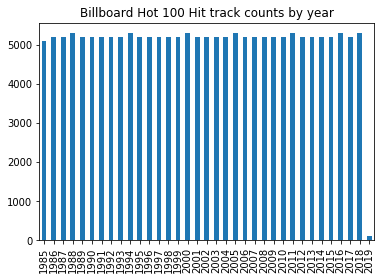}
        \caption{Billboard track counts by year. Years with few than 5k songs were excluded from our dataset (e.g. 2019).}
        \centering
        \label{fig:billboard_track_counts}
    \end{subfigure}
    \begin{subfigure}{0.5\textwidth}
        \includegraphics[width=7cm, left]{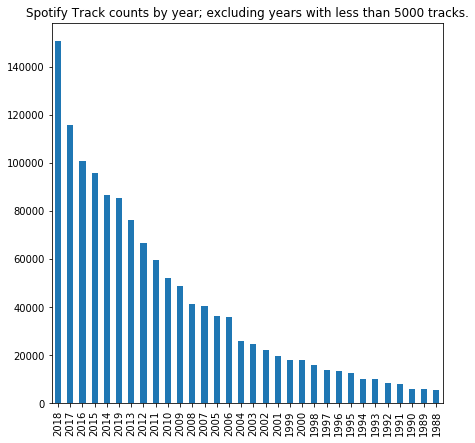}
        \caption{Spotify track counts by year. Years with few than 5k songs were excluded from our dataset.}
        \centering
        \label{fig:spotify_track_counts}
    \end{subfigure}
    \caption{Track counts by year in both the Spotify and Billboard datasets}
    \label{fig:dataset_distributions}
\end{figure}

Previous work on HSS have used relatively small datasets \cite{hitpredict}. We extend previous work by creating a larger dataset. We believe this larger dataset will allow for more robust model architectures than previous datasets. We used the Spotify API to create a dataset with approximately 1.8 million songs. We reduced the size of the dataset by considering only the songs released between the years 1985 and 2018. We then collected a dataset of all unique songs on the Billboard Hot 100 chart between 1985 and 2018 using the Billboard API ($\sim$16k songs). See figure \ref{fig:billboard_track_counts}. Finally, we merged the Spotify and Billboard datasets together by matching tracks on their title and artist. This combined dataset contained approximately 1.8 million tracks, with approximately 12,000 tracks being Billboard Hot 100 hits. See figure \ref{fig:spotify_track_counts}.

In order to balance our data, we randomly sampled 12,000 non-hits from the Spotify data and created a new dataset. This dataset contained approximately 12k non-hits and 12k hits ($\sim$24k tracks total). We refer to this dataset as SpotifyBillboard.

Each track contains 27 features categorized by track information, artist information, album information, and audio analysis features. We describe these feature in detail below:
\begin{itemize}
    \item \textit{track\_id}: the song's unique Spotify track ID 
    \item \textit{track\_title}: the track title
    \item \textit{artist\_title}: the artist's title
    \item \textit{artist\_id}: the artist's unique Spotify ID
    \item  \textit{popularity}: a value between 0 and 100, with 100 being the most popular.  Popularity is calculated by Spotify, and is based, "in the most part, on the total number of plays the track has had and how recent those plays are" \cite{spotify_for_developers_2019}.
    \item \textit{explicit}: a value indicated whether a track has explicit lyrics (1 = explicit, 0 = not explicit)
    \item  \textit{duration\_ms}: the duration of the track in milliseconds. 
    \item \textit{preview\_url}: a link to a 30 second preview of the track.
    \item \textit{album\_id}: the unique Spotify album ID
    \item \textit{album\_type}: the type of the album: one of “album”, “single”, or “compilation”
    \item \textit{album\_release\_date}: the date the album was first released (date format: YYYY-MM-DD)
    \item \textit{acousticness}: a value from 0.0 to 1.0 predicting whether the track is acoustic.
    \item \textit{danceability}: a value from 0.0 to 1.0 describing how suitable a track is for dancing based on a combination of musical elements including tempo, rhythm stability, beat strength, and overall regularity.  Values closer to 1.0 indicate that the track is more danceable.
    \item \textit{energy}:  a value from 0.0 to 1.0 that represents a perceptual measure of intensity and activity
    \item \textit{instrumentalness}: a value from 0.0 to 1.0 predicting whether the track is instrumental or contains vocals. Values closer to 1.0 represent more instrumental track. 
    \item \textit{key}: the key the track is in
    \item \textit{liveness}: a value from 0.0 to 1.0 that describes the presence of an audience in the track. Values closer to 1.0 represent tracks that were performed live. 
    \item\textit{loudness}: the overall loudness of a track in decibels (dB)
    \item\textit{mode}: indicates the modality (major=1 or minor=0) of a track.
    \item \textit{speechiness}: a value from 0.0 to 1.0 describing the amount of spoken words present in the track. Values close to 1.0 indicate exclusively speech-like tracks (e.g. podcast, audio book, poetry).
    \item \textit{tempo}: the overall estimated tempo of a track in beats per minute (BPM)
    \item \textit{time\_signature}: an estimated overall time signature of a track.
    \item \textit{valence}: a value from 0.0 to 1.0 describing the musical positiveness conveyed by a track. A value close to 1.0 suggests that the track sounds more positive and upbeat. 
    \item \textit{weeks}: a value indicating the total number of weeks the track was on the Billboard Hot 100 chart
    \item \textit{rank}: a value between 0 and 100 indicating a track's position on the Billboard Hot 100 chart. A value of 0 indicates that the track never appeared on the Billboard Hot 100 chart.
    \item \textit{score}: a weighted rank value from 0.0 to 1.0 indicating the popularity of a track. The score is a custom made value. It is a weighted value indicating the most popular tracks on the Billboard Hot 100 chart. A value of 0.0 indicates the track never appeared on the chart. A value close to 1.0 indicates that the track appeared frequently at the of the chart. Note, this value was not given to us, we used a data mining method, which was described above, to calculate this value. 
    \item \textit{billboard\_hit}: indicates whether the track appeared on the Billboard Hot 100 chart (1=hit, 0=non-hit).
\end{itemize}

\begin{figure}[ht]
    \centering
    \includegraphics[width=0.5\textwidth]{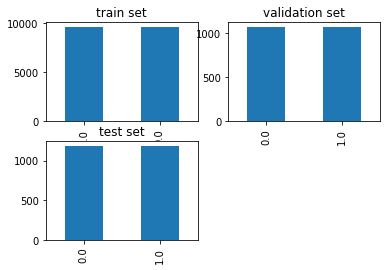}
    \caption{Distribution of Hits and Non-Hits in the train (20k tracks), validation (2k tracks), and test (2k tracks) sets.}
    \centering
    \label{fig:train_val_test_dist}
\end{figure}

After processing SpotifyBillboard features we created the train, validation, and test sets. No tracks in the training set appeared in the validation and test set. Each set contained a balanced distribution between hits and non-hits (Figure \ref{fig:train_val_test_dist}). Note, the majority of HSS work has only considered audio analysis features (acousticness, instrumentalness, etc.) \cite{hitpredict, herremans2014dance}. To extend previous work, in addition to audio analysis features, we consider song duration and mine an additional artist past-performance feature. Artist past-performance for a given song represents how many prior Billboard hits the artist has released before that track's release date. 

\subsection{Training Environment}
We utilized the entire fleet of computers available to University of San Francisco's Computer Science Department ($\sim$40) in order to run first a randomized search and then a grid search on each of the classification algorithms available in the scikit-learn package. The average training time was about one day and the following models have been reported for notable performance. 

\subsection{Models \& Algorithms}
To predict whether a song will be a Billboard hit or not, we use four different models:
\begin{itemize}
    \item \textit{Logistic Regression (LR)}
    \item \textit{Neural Network (NN)}
    \item \textit{Random Forest (RF)}
    \item \textit{Support Vector Machine (SVM)}
\end{itemize}

\textbf{Logistic regression} (LR) is a popular classification algorithm. It is used when the dependent (target) variable is categorical. The idea in LR is to find a relationship between features and the probability of a particular outcome. There are two types LR problems binary logistic regression and multi-class logistic regression. We used binary logistic regression because our dependent variable has two possible values 0 (non-hit) and 1 (hit). We use the sigmoid activation function to constrain our probability estimate between 0 and 1.
\begin{equation}\label{sigmoid}
\sigma(x) = \frac{e^{x}}{1+e^{x}}
\end{equation}
We use Maximum Likelihood Estimation (MLE) to estimate the feature coefficients and RMSEprop to back-propagate the gradients over 1000 epochs. We define the cost function below.

\begin{equation} \label{MLE_LR}
\begin{split}
    L(\beta;y) & = \prod_{i=1}^n P\big(Y_i=y_i | X_i=x_i\big)\\
               & = \prod_{i=1}^n \sigma\big(x_i^{t}\beta\big)^{y_i} \big(1-\sigma\big(x_i^{t}\beta\big)\big)^{1-y_i}
\end{split}  
\end{equation}

Where $\sigma\big(x_i^{t}\beta\big)$ is the probability of a hit and  
$\big(1-\sigma\big(x_i^{t}\beta\big)\big)$ is the probability of a non-hit. Additionally, $y_i$ = $1$ (hit) or $0$ (non-hit). 

\begin{figure}[ht]
    \centering
    \includegraphics[width=0.5\textwidth]{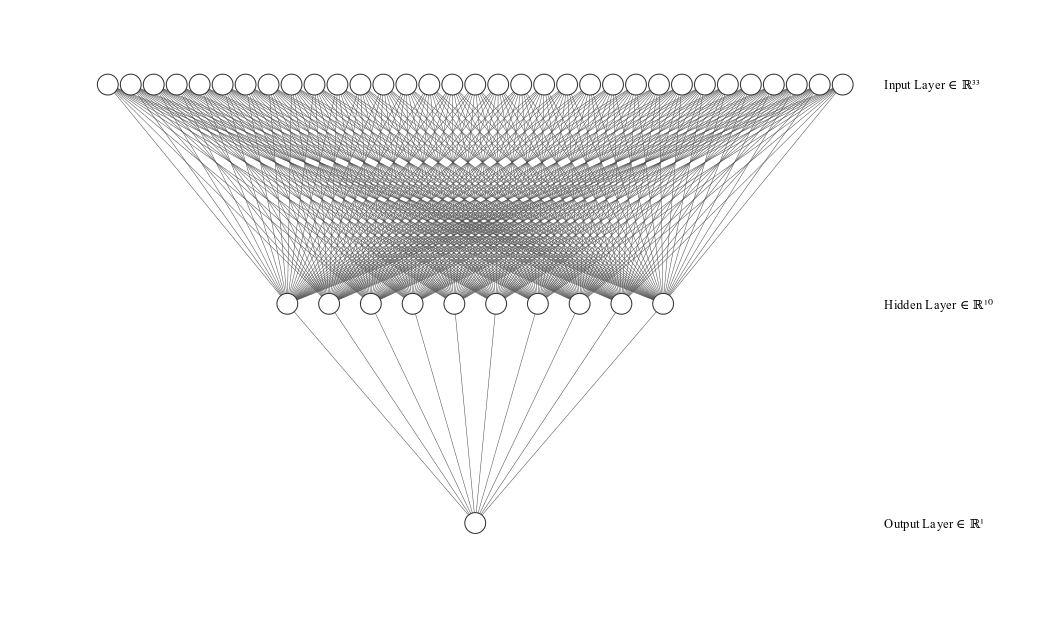}
    \caption{Diagram of our neural network architecture. We use 1 hidden layer with 10 filters.}
    \centering
    \label{fig:NN_diagram}
\end{figure}

\textbf{Neural Networks} (NNs) have become popular to solve classification tasks after the rise of deep learning. We use a simple neural network with one hidden layer to solve HSS (Figure \ref{fig:NN_diagram}). We use RMSprop-an unpublished optimization algorithm designed for neural networks, first proposed by Geoff Hinton \cite{RMSEprop}, and sigmoid function in the final layer to constrain the output between 0 and 1. In the hidden layer, we use ten filters and rectified linear unit (ReLU) activation. We set the batch size to 32 and stopped training after 1000 epochs. See figure \ref{fig:NN_diagram}.

\textbf{Random Forest} (RF) models are one of the most popular ensemble methods used in classification. These models aim to correct for the problem of over-fitting in traditional decision trees. This will not be covered in depth, but decision trees tend to learn on irregular paths of data. RF models train multiple deep decision trees on different aspects of the dataset with the aim of reducing the overall variance. Not only was RF the most accurate model overall, but it was the quickest to train. We used a maximum number of features of eight with 80 estimators and a minimum split condition of two samples under the Gini criterion. 

\textbf{Support Vector Machine} (SVM) aims to find the most optimal hyper-plane that separates the data into two distinct classes. We used the Gaussian Radial Basis Function (RBF) as our kernel: $\exp(-\gamma \|x-x'\|^2)$. Our model uses $\gamma = 0.1$ and $C = 10$. 

\section{Results}
\label{sec:results}
We focused mainly on the accuracy of results, but we report the precision and recall as well since false positive predictions may be costly when a music label invests in a song that is actually unlikely to become a hit (Table \ref{tab:model_results}).

\begin{table}[ht]
    \caption{Model Results}
    \label{tab:model_results}
    \centering
    \resizebox{0.75\columnwidth}{!}{
    \begin{tabular}{|l|l|l|l|l|l|l|}
        \hline
        \multirow{2}{*}{Models} & \multicolumn{2}{l|}{Accuracy} & \multicolumn{2}{l|}{Precision} & \multicolumn{2}{l|}{Recall} \\ \cline{2-7} 
        & Test & Val & Test & Val & Test & Val \\ \hline
        Logistic Regression & 0.8151 & 0.8065 & 0.7526 & 0.7457 & 0.9391 & 0.9298 \\ \hline
        Neural Network & 0.8214 & 0.8305 & 0.8235 & 0.8233 & 0.7913 & 0.7671 \\ \hline
        \textbf{Random Forest} & \textbf{0.877} & \textbf{0.887} & \textbf{0.86} & \textbf{0.87} & \textbf{0.9} & \textbf{0.89}  \\ \hline
        SVM & 0.839 & 0.828 & 0.995 & 0.993 & 0.704 & 0.706 \\ \hline
    \end{tabular}
    }
\end{table}
\vline

The NN model with one hidden layer gave 82.14\% and 83.05\% accuracy on the validation and test data, with similar results on the training data indicating no over-fitting. The final cross-entropy loss after 1000 epochs was 0.4261. The precision and recall  on the validation set were 82.33\% and 76.71\%.  The confusion matrix on the validation set shows that there are some false positives (Table \ref{tab:NN_confusion_matrix}).
\begin{table}[h!]
    \centering
    \caption{NN Confusion Matrix on the validation set}
    \label{tab:NN_confusion_matrix}
    \begin{tabular}{cc|cc}
        \multicolumn{1}{c}{} &\multicolumn{1}{c}{} &\multicolumn{2}{c}{Actual} \\ 
        \multicolumn{1}{c}{} & 
        \multicolumn{1}{c|}{} & 
        \multicolumn{1}{c}{Hit} & 
        \multicolumn{1}{c}{Non-Hit} \\ \hline
        \multirow{2}{*}{\rotatebox{90}{Predicted}}
        & Hit  & 1027 & 363 \\[1.5ex]
        & Non-Hit  & 42  & 707 \\ 
    \end{tabular}
    \quad
\end{table}

The LR model yielded 80.65\% accuracy on the validation data and 81.51\% accuracy on the test data, with similar result on the training data indicating no over-fitting. The precision and recall were acceptable at 74.57\% and 92.98\%. The confusion matrix on the validation set shows that there are some false positives (Table \ref{tab:LR_confusion_matrix}). 
\begin{table}[h!]
    \centering
    \caption{LR Confusion Matrix on the validation set}
    \label{tab:LR_confusion_matrix}
    \begin{tabular}{cc|cc}
        \multicolumn{1}{c}{} &\multicolumn{1}{c}{} &\multicolumn{2}{c}{Actual} \\ 
        \multicolumn{1}{c}{} & 
        \multicolumn{1}{c|}{} & 
        \multicolumn{1}{c}{Hit} & 
        \multicolumn{1}{c}{Non-Hit} \\ \hline
        \multirow{2}{*}{\rotatebox{90}{Predicted}}
        & Hit  & 994 & 339 \\[1.5ex]
        & Non-Hit  & 75  & 731 \\
    \end{tabular}
    \quad
\end{table}

The RF model yielded 88.7\% accuracy on the validation data and 87.7\% accuracy on the test data, with similar result on the training data indicating no over-fitting. The precision and recall were acceptable at 87\% and 89\%. The confusion matrix on the validation set shows that there are some false positives and false negatives (Table \ref{tab:RF_confusion_matrix}). 
\begin{table}[h!]
    \centering
    \caption{RF Confusion Matrix on the validation set}
    \label{tab:RF_confusion_matrix}
    \begin{tabular}{cc|cc}
        \multicolumn{1}{c}{} &\multicolumn{1}{c}{} &\multicolumn{2}{c}{Actual} \\ 
        \multicolumn{1}{c}{} & 
        \multicolumn{1}{c|}{} & 
        \multicolumn{1}{c}{Hit} & 
        \multicolumn{1}{c}{Non-Hit} \\ \hline
        \multirow{2}{*}{\rotatebox{90}{Predicted}}
        & Hit  & 917 & 153 \\[1.5ex]
        & Non-Hit  & 82  & 993 \\ 
    \end{tabular}
    \quad
\end{table}

The SVM model yielded 82.8\% accuracy on the validation data and 83.9\% accuracy on the test data, with similar result on the training data indicating no over-fitting. The precision and recall were acceptable at 79\% and 89\%. The confusion matrix on the validation set shows that there are some false negatives (Table \ref{tab:SVM_confusion_matrix}).
\begin{table}[h!]
    \centering
    \caption{SVM Confusion Matrix on the validation set}
    \label{tab:SVM_confusion_matrix}
    \begin{tabular}{cc|cc}
        \multicolumn{1}{c}{} &\multicolumn{1}{c}{} &\multicolumn{2}{c}{Actual} \\ 
        \multicolumn{1}{c}{} & 
        \multicolumn{1}{c|}{} & 
        \multicolumn{1}{c}{Hit} & 
        \multicolumn{1}{c}{Non-Hit} \\ \hline
        \multirow{2}{*}{\rotatebox{90}{Predicted}}
        & Hit  & 1065 & 5 \\[1.5ex]
        & Non-Hit  & 447  & 628 \\ 
    \end{tabular}
    \quad
\end{table}

\section{Conclusion \& Future Work}
\begin{figure}[h]
    \centering
    \includegraphics[width=0.5\textwidth]{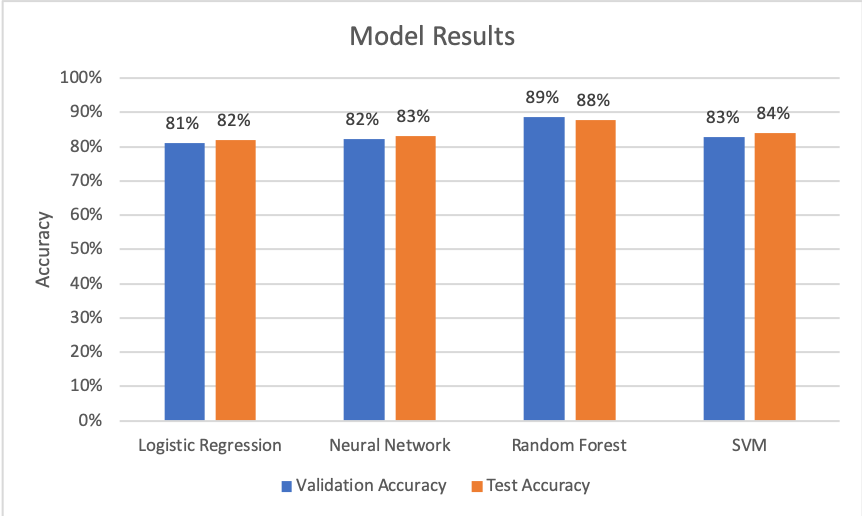}
    \caption{Model Results on the validation and test sets. The Random Forest model was the most successful}
    \label{fig:model_results}
\end{figure}

The results showed that SVM and RF outperform LR and NN in regards to accuracy (Figure \ref{tab:model_results}). The most robust model is the RF. Interestingly, the SVM had the highest precision accuracy (Table \ref{tab:model_results}). 

The false positive rate for our SVM is very low, while maintaining an average false negative rate. The truth values predicted by this model can be trusted while the false values cannot. This algorithm is greedy and will assume the least amount of risk when classifying a positive. Music labels may prefer to use the SVM since it is less likely to predict hits incorrectly.

In future experiments we would like to investigate label influence and social media presence with respect to song success. Using features of the audio itself combined with artist past-performance has managed to explain a majority of the variance in the data; we believe there are more types of features which can provide our model with a social context to make even better predictions.

\bibliographystyle{unsrt}  
\bibliography{song_hit_prediction}  






\end{document}